\documentclass{ws-procs9x6}

\begin{document}

\title{Dark energy, dark matter and 
\\
fermion families in the
two measures theory}

\author{ E. I. Guendelman and A.  B.  Kaganovich }

\address{ Physics Department,
\\
 Ben Gurion University of the Negev,
\\
 Beer Sheva 84105, Israel
 \\ 
E-mail: guendel@bgumail.bgu.ac.il  and alexk@bgumail.bgu.ac.il }

\maketitle

\abstracts{A field theory is proposed where the regular fermionic matter 
and the dark fermionic matter are different states of the same "primordial" fermion fields.
In regime of the fermion densities typical for normal particle physics, 
each of the primordial fermions splits into three generations identified with regular fermions. 
In a simple model, this  fermion families 
birth effect is accompanied with the right lepton numbers conservation
laws. It is possible to fit the muon to electron mass ratio without fine
tuning of the Yukawa coupling constants.
When fermion energy density becomes comparable with dark energy density, 
the theory allows new type of states - Cosmo-Low Energy Physics
(CLEP) states. Neutrinos in CLEP state can be both a good candidate for 
dark matter and responsible for a new type of dark energy.
In the latter case the total energy density of the universe is less than it would be in
the universe free of fermionic matter at all. The (quintessence) scalar field
is coupled to dark matter but its coupling to regular fermionic matter
appears to be extremely suppressed.}

\section{Introduction}

The existence of the lepton and quark generations is one of the greatest
puzzles of the particle physics. Resolution of this puzzle implies an explanation 
of the origin of the mass spectrum of the elementary fermions and observable flavor
properties of their electroweak interaction. In this talk side by side with the
fermion families problem we will study the dark matter and dark energy puzzle. It
turns out that these, for the first glance, absolutely different  problems are 
very much connected in the framework
of the so called Two Measures Theory (TMT).

Understanding of the nature of the dark matter and dark energy should
be the basis for explanation of the cosmic coincidence. A promising 
approach towards solving the cosmic
coincidence problem in the scenario of an accelerating expansion
for the present day universe was developed in the variable mass
particles (VAMP) models\cite{VAMP}. However such a modification of the
 particle physics theory underlying the quintessence scenarios
has a fundamental problem:
 although there are  some  justifications for choices of
certain types of dark matter-dark energy coupling  in the Lagrangian,
there is a necessity to assume
the absence or extremely strong suppression of the barion matter-dark energy   
coupling. Actually this problem was known from the
very beginning  in the
quintessence models since generically there are no reasons for the absence of  a
direct coupling of the quintessence scalar field $\phi$ to the barion matter.
Such coupling  would be the origin of a long range scalar force because
of the very small mass of the quintessence field  $\phi$.
This "fifth-force" problem
might be solved\cite{Carroll} if there would be a shift symmetry
$\phi\rightarrow\phi +const$ of the action.
 However the quintessence potential itself does not
possess this symmetry. The situation with the fifth-force problem becomes
still more critical in the discussed above models since one should explain now
why the direct quintessence-dark matter coupling is permissible in the Lagrangian
while the same
is forbidden for the barion matter.

Modifications of the particle physics models in
Refs.\cite{VAMP}
are based on
 the assumption
that   all the fields of  the fundamental particle theory should be divided into two
large groups:  one describing  detectable particles (ordinary matter) and the other
including dark matter particles.
The main purpose of this talk is to point out that TMT 
is able to propose a resolution for the above problems by an absolutely new way:
{\em the dark matter is not introduced as a special type of matter but rather it
appears as the solution of equations of motion describing a new type of states}
 of the (primordial) neutrino field; in other
words, {\em  the dark neutrino matter and the regular neutrino generations
 (electron, muon
and $\tau$ neutrinos) are different states of the same primordial fermion
field. }.

TMT has been originally constructed
with
the aim to solve
the "old" cosmological constant problem\cite{GK1}. TMT is a generally
coordinate invariant theory
with the action of the following general form
\begin{equation}
    S = \int L_{1}\Phi d^{4}x +\int L_{2}\sqrt{-g}d^{4}x
\label{S}
\end{equation}
where $\Phi$ is a scalar density built of four scalar
fields
$\varphi_{a}$ ($a=1,2,3,4$)
\begin{equation}
\Phi
=\varepsilon^{\mu\nu\alpha\beta}\varepsilon_{abcd}\partial_{\mu}\varphi_{a}
\partial_{\nu}\varphi_{b}\partial_{\alpha}\varphi_{c}
\partial_{\beta}\varphi_{d}
\label{Phi}
\end{equation}
and $L_{1}$, $L_{2}$ are independent of the measure fields
$\varphi_{a}$.

We proceed in the first order formalism
where  all fields, including
also vierbeins ${e}_{a\mu}$, spin-connection $\omega_{\mu}^{ab}$ and the
measure fields $\varphi_{a}$ are independent dynamical variables. All the
relations between them are results of equations of motion.
It turns out that the measure
fields $\varphi_{a}$ affect the
theory only via
the scalar field $\zeta\equiv \Phi /\sqrt{-g}$ which is determined by
a constraint in the form of an algebraic equation. The latter is exactly
a consistency condition of
equations of motion and {\it determines $\zeta$ in terms of matter fields}.
 After  transformation to new variables
(conformal Einstein frame),
all equations of motion take canonical GR form of equations for   
gravity and matter fields.
 All the novelty consists in the  structure of the dilaton $\phi$ and
the Higgs field effective potentials, masses and
interactions of fermions
as well as the structure of their contributions to the energy-momentum
tensor: all these now depend on the fermion densities via $\zeta$.

\section{$SU(2)\times U(1)$ gauge invariant and scale-invariant model}

The
TMT model
we have studied in Ref.\cite{nnu} possesses spontaneously broken
global
scale symmetry\cite{G1} which
allows to suggest\cite{GK4} a simultaneous resolution of both the
fermion families problem and the fifth-force problem. {\em The theory
starts from one primordial fermion field for each type of leptons and
quarks}. For example, in $SU(2)\times U(1)$ gauge theory the fermion content is the
left doublets and right singlets constructed from the  primordial neutrino $N$
 and electron $E$ fields 
and primordial $U$ and $D$ quark fields. 

We allow in both $L_{1}$ and $L_{2}$ all the usual contributions
considered in standard field theory models in curved space-time. 
Keeping the general structure
(\ref{S}), it is convenient to represent the integrand of the
 action in the following 
form:
\begin{eqnarray}
&&\Phi L_{1}+\sqrt{-g}L_{2}
\nonumber\\
&&= e^{\alpha\phi /M_{p}}
(\Phi +b\sqrt{-g})\left[-\frac{1}{\kappa}R(\omega ,e)
+\frac{1}{2}g^{\mu\nu}\phi_{,\mu}\phi_{,\nu}+
\frac{1}{2}g^{\mu\nu}(D_{\mu}H)^{\dag}D_{\nu}H\right]
\nonumber\\   
&&-e^{2\alpha\phi /M_{p}}[\Phi V_{1}(H) +\sqrt{-g}V_{2}(H)]
+e^{\alpha\phi /M_{p}}(\Phi +k\sqrt{-g})L_{fk}
\nonumber\\
     &&-e^{\frac{3}{2}\alpha\phi /M_{p}}
\left[(\Phi +h_{E}\sqrt{-g})f_{E}\overline{L}_{L}\,H\,E_{R}
+h.c.\right]
\nonumber\\
&&-e^{\frac{3}{2}\alpha\phi /M_{p}}
\left[(\Phi +h_{N}\sqrt{-g})f_{N}\overline{L}_{L}\,H^{c}\,N_{R}
+h.c.\right]
\nonumber\\
&&-\frac{1}{4}g^{\alpha\mu}g^{\beta\nu}
\left(B_{\alpha\beta}B_{\mu\nu}+W^{a}_{\alpha\beta}W^{a}_{\mu\nu}\right)
\label{totaction}
\end{eqnarray}
Here $R(\omega ,e)$ is the scalar curvature in the vierbein - spin-connection
formalism;
 $L_{fk}$ is the fermion kinetic term including the standard interaction
to the gauge fields $\vec{W}_{\mu}$ and $B_{\mu}$; $H$ is the Higgs
doublet. Constants $b, k, h_{N}, h_{E}$ are non specified dimensionless real parameters
of the model and we will only assume that {\em  orders of their magnitudes are not
too much different}. The real positive parameter $\alpha$ is
assumed to be of the order of one. For short we omit here and in what follows
the primordial quarks since they appear in a form similar to the primordial
leptons. 
The action incorporates
also the dilaton field $\phi$ which provides global scale symmetry
($\Psi$ denotes all the primordial fermions):
\begin{eqnarray}
&& e_{\mu}^{a}\rightarrow e^{\theta /2}e_{\mu}^{a}, \quad
\omega^{\mu}_{ab}\rightarrow \omega^{\mu}_{ab}, \quad
\varphi_{a}\rightarrow \lambda_{a}\varphi_{a}\quad
where \quad \Pi\lambda_{a}=e^{2\theta},
\nonumber
\\
&&\phi\rightarrow \phi-\frac{M_{p}}{\alpha}\theta ,\quad
\Psi\rightarrow e^{-\theta /4}\Psi, \quad
\overline{\Psi}\rightarrow  e^{-\theta /4} \overline{\Psi};
\quad \theta =const,
\nonumber
\\
&&H\rightarrow H , \quad \vec{W}_{\mu}\rightarrow \vec{W}_{\mu}  \quad B_{\mu}\rightarrow B_{\mu}.
\label{stferm}
\end{eqnarray}
 In the
context of cosmology, $\phi$ plays the role of a quintessence
field.

In the new variables
($\phi$,
$H$, $B_{\mu}$ and $\vec{W}_{\mu}$ remain unchanged)
which we call 
the Einstein frame,
\begin{eqnarray}
\tilde{g}_{\mu\nu}&=&e^{\alpha\phi/M_{p}}(\zeta +b)g_{\mu\nu}, \quad
\tilde{e}_{a\mu}=e^{\frac{1}{2}\alpha\phi/M_{p}}(\zeta
+b)^{1/2}e_{a\mu}, 
\nonumber\\
\Psi^{\prime}_{i}&=&e^{-\frac{1}{4}\alpha\phi/M_{p}}
\frac{(\zeta +k)^{1/2}}{(\zeta +b)^{3/4}}\Psi_{i} , \quad
i=N,E,
\label{ctferm}
\end{eqnarray}
the gravitational equations take the form
$G_{\mu\nu}(\tilde{g}_{\alpha\beta})=\frac{\kappa}{2}T_{\mu\nu}^{eff}$
where $G_{\mu\nu}(\tilde{g}_{\alpha\beta})$ is the Einstein tensor
in the Riemannian space-time with the metric
$\tilde{g}_{\mu\nu}$ and
\begin{eqnarray}
&&T_{\mu\nu}^{eff}=\phi_{,\mu}\phi_{,\nu}-\frac{1}{2}
\tilde{g}_{\mu\nu}\tilde{g}^{\alpha\beta}\phi_{,\alpha}\phi_{,\beta}
+(D_{\mu}H)^{\dag}D_{\nu}H
-\frac{1}{2}\tilde{g}_{\mu\nu}\tilde{g}^{\alpha\beta}(D_{\alpha}H)^{\dag}D_{\beta}H
\nonumber\\
&+&\tilde{g}_{\mu\nu}V_{eff}(\phi,\upsilon ;\zeta)
+T_{\mu\nu}^{(gauge)}
+T_{\mu\nu}^{(ferm,can)}+T_{\mu\nu}^{(ferm,noncan)};
 \label{Tmn}   
\end{eqnarray}
\begin{equation}
V_{eff}(\phi,\upsilon ;\zeta)=
\frac{b\left[M^{4}e^{-2\alpha\phi/M_{p}}+V_{1}(\upsilon )\right]
-V_{2}(\upsilon)}{(\zeta
+b)^{2}};
\label{Veff1}
\end{equation}
$T_{\mu\nu}^{(gauge)}$ is the canonical energy momentum tensor for the
$SU(2)\times U(1)$ gauge fields sector;
$T_{\mu\nu}^{(ferm,can)}$ is the canonical energy momentum tensor for
(primordial) fermions $N^{\prime}$ and $E^{\prime}$ in
curved space-time
including also their standard $SU(2)\times U(1)$ gauge interactions.
$T_{\mu\nu}^{(ferm,noncan)}$ is the {\em noncanonical} contribution
of the fermions into the energy momentum tensor
$ T_{\mu\nu}^{(ferm,noncan)}=-\tilde{g}_{\mu\nu}[F_{N}(\zeta,\upsilon)
\overline{N^{\prime}}N^{\prime}+
F_{E}(\zeta,\upsilon)\overline{E^{\prime}}E^{\prime}]\quad\equiv\quad
-\tilde{g}_{\mu\nu}\Lambda_{dyn}^{(ferm)}$. The scalar field $\zeta$
is determined by the constraint
\begin{equation}
\frac{1}{(\zeta +b)^{2}}\left\{(b-\zeta)\left[M^{4}e^{-2\alpha\phi/M_{p}}+
V_{1}(\upsilon)\right]-2V_{2}(\upsilon)\right\} 
=\Lambda_{dyn}^{(ferm)}
\label{constraint3}
\end{equation} 
where
\begin{equation}
F_{i}(\zeta,\upsilon)\equiv
\frac{\upsilon f_{i}}{2\sqrt{2}(\zeta +k)^{2}(\zeta +b)^{1/2}}
(\zeta -\zeta^{(i)}_{1})(\zeta -\zeta^{(i)}_{2}),
\quad i=N,E,
 \label{Fizeta1}
\end{equation}
\begin{equation}
\zeta_{1,2}^{(i)}=\frac{1}{2}\left[k-3h_{i}\pm\sqrt{(k-3h_{i})^{2}+
8b(k-h_{i})
-4kh_{i}}\,\right], \quad i=N,E.
\label{zetapm}
\end{equation}

The $\zeta$ depending "masses" of the primordial fermions
are of the form
\begin{equation}
m_{N}(\zeta,\upsilon)=
\frac{\upsilon f_{N}(\zeta^{(N)} +h_{N})}{\sqrt{2}(\zeta^{(N)} +k)(\zeta^{(N)} +b)^{1/2}},
\,
m_{E}(\zeta,\upsilon)=
\frac{\upsilon f_{E}(\zeta^{(E)} +h_{E})}{\sqrt{2}(\zeta^{(E)} +k)(\zeta^{(E)} +b)^{1/2}}
 \label{muferm1}
\end{equation}
where $\upsilon$ is the VEV of the Higgs field in the unitary gauge.

\section{Fermions in normal particle physics conditions}

The simple analyze shows that at fermion energy densities corresponding
to  {\it normal laboratory particle physics} conditions
("high fermion density") the
balance  imposed by the constraint can be realized only through one of the following
 two ways:
(I) $F_{i}(\zeta,\upsilon)\approx 0, \, \Rightarrow \,
\zeta =\zeta_{1}^{(i)} \, or \, \zeta =\zeta_{2}^{(i)},  \, i=N,E$;
\\
(II) $\zeta =\zeta_{3}\approx -b$.

Two constant solutions $\zeta^{(i)}_{1,2}$ correspond
to two different states of the primordial leptons with {\it
different constant
masses} determined by Eq.(\ref{muferm1}) where we have to substitute
$\zeta_{1,2}^{(i)}$ instead of $\zeta$.
We {\it identify these two states of the primordial leptons with the mass
 eigenstates of  the first two generations of the
regular leptons}. If the free primordial electron is in the state with $\zeta =
\zeta^{(E)}_{1}$ (or $\zeta =\zeta^{(E)}_{2}$) it is detected as the regular
electron  $e$ (or muon  $\mu$) and similar for the electron and muon neutrinos
with masses respectively:
\begin{equation}
m_{\nu_{e}(\nu_{\mu})}= \frac{\upsilon_{0}f_{N} (\zeta^{(N)}_{1(2)}
+h_{N})}
{\sqrt{2} (\zeta^{(N)}_{1(2)}  +k)(\zeta^{(N)}_{1(2)}  +b)^{1/2}};
\,
m_{e(\mu)}= \frac{\upsilon_{0}f_{E} (\zeta^{(E)}_{1(2)} +h_{E})}
{\sqrt{2} (\zeta^{(E)}_{1(2)}  +k)(\zeta^{(E)}_{1(2)}  +b)^{1/2}}
\label{m-E}
\end{equation}
One can show that the model provides right flavour properties
of the electroweak interactions, at least for the first two
lepton generations.

Turning now to the old problem of the ratio
$m_{\mu}/m_{e}\approx 207$ one can notice at once that the  fitting of the
Yukawa couplings has no relation to this
problem because electron and muon emerge as different states of the same primordial field $E^{\prime}$.
However, our TMT model opens a possibility to get the desirable ratio $m_{\mu}/m_{e}$
by means of a soft enough restrictions on the dimensionless   
parameters $b$, $k$ and $h_{E}$. We will assume in what follows that
\begin{equation}
b>0, \quad k<0, \quad h_{E}<0, \quad \frac{|k|}{b}=n>2 \quad and \quad |h_{E}|\gg|k|
\label{parameters}
\end{equation}
Then it follows from Eq.(\ref{zetapm}) that
$\zeta_{1}\approx 3|h_{E}|-3n(4n-2)/|k|, \quad
 \zeta_{2}\approx 3^{-1}|k|(1-2/3n)$
 and Eq.(\ref{m-E}) gives $m_{\mu}/m_{e}\approx 207$ if
\begin{equation}
\frac{|h_{E}|}{|k|}=9.8\frac{n+1}{n}
\label{par-for-m-e}
\end{equation}
One can show that our TMT model provides the universality, 
(i.e. $\zeta$ independence)
of the Lagrangian of the electro-weak interaction. Therefore
with the same condition (\ref{par-for-m-e}) we obtain the right ratio
 of the physical (renormalized) muon and electron  masses.

 The effective  interaction of the dilaton $\phi$
with the regular leptons and quarks of the first two generations appears
to be extremely suppressed because the appropriate Yukawa coupling 
is proportional to $F_{i}(\zeta,\upsilon)$.
 In other words, the interaction of the dilaton with matter observable
 in  gravitational experiments is
 practically switched off, and that solves the
fifth-force problem.

The solution of the type II, $\zeta^{(i)} =\zeta_{3}^{(i)}\approx -b$
 we associate with
the third fermion generations. These  states should be realized
via fermion condensate.

\section{Vacuum and/or very low fermion
density?}

One can show that in the fermion vacuum, the effective potential
of the scalar sector including the quintessence-like field $\phi$ and 
the Higgs field $\upsilon$ is 
\begin{equation}
V_{eff}^{(0)}(\phi)
=\frac{[V_{1}(\upsilon_{0})+M^{4}e^{-2\alpha\phi/M_{p}}]^{2}}
{4[b\left(V_{1}(\upsilon_{0})+M^{4}e^{-2\alpha\phi/M_{p}}\right)-V_{2}(\upsilon_{0})]}
\label{Veffvac}
\end{equation}
where $\upsilon_{0}$ is determined by the equation 
\begin{equation}
\left[b-2V_{2}(\upsilon_{0})
\left(V_{1}(\upsilon_{0})+M^{4}e^{-2\alpha\phi/M_{p}}\right)^{-1}\right]
V_{1}^{\prime}(\upsilon_{0})
+V_{2}^{\prime}(\upsilon_{0})=0
\label{eq-for-zeta-0}
\end{equation}

The structure of the potential (\ref{Veffvac}) allows to construct a model,
where zero vacuum energy is achieved without fine tuning\cite{GK1} when  
$V_{1}(\upsilon_{0})+M^{4}e^{-2\alpha\phi/M_{p}}=0$. In fact one may get such situation multiple times, therefore 
naturally obtaining 
a multiple degenerate vacuum as advocated by Bennett, Froggatt and Nielsen\cite{Nilsen}.

In alternative models where the potential $V_{eff}^{(0)}(\phi)$ has no zeros, it can
monotonically decrease to the cosmological constant
\begin{equation}
\Lambda^{(0)}
=\frac{V_{1}^{2}(\upsilon_{0})}
{4[bV_{1}(\upsilon_{0})-V_{2}(\upsilon_{0})]}.
\label{lambda-without-ferm}
\end{equation}
This takes place if $bV_{1}(\upsilon_{0})>2V_{2}(\upsilon_{0})$. Then in the context of cosmology of the late
time universe, $\phi$ plays the role of the quintessence-like field and (\ref{Veffvac}) is
the dark energy potential. 

Physics of fermions at very low densities, as it is governed by the
constraint (\ref{constraint3}), turns out to be very different
from what we know in normal particle physics. The term {\it very low fermion
density} means here that the fermion
energy density is comparable with the dark energy density. In this
case, the  noncanonical contribution 
(proportional to $ F_{i}(\zeta,\upsilon)\bar{\Psi}_{i}\Psi_{i}$) to the  energy-momentum
tensor of the primordial fermion fields can be larger and even much larger
 than the canonical one.
 The theory predicts that in
this regime the
primordial fermion
may not split into  generations. Instead of this, for instance, in the FRW universe,
the
primordial fermion can
 participate in the expansion of the
universe by means of changing its own parameters. We call this effect
"Cosmo-Particle Phenomenon"  and refer to such states
as  Cosmo-Low Energy Physics (CLEP) states.

As the first step in studying Cosmo-Particle Phenomena,
we restrict ourselves to the consideration of a simplified cosmological
model where universe  is filled with
a homogeneous scalar field $\phi$ and uniformly distributed
{\it non-relativistic (primordial) neutrinos} in CLEP states.
A possible way to get up such a CLEP state might be
spreading of the wave packet during its free motion lasting
a very long (of the cosmological scale) time.
The constraint shows that decreasing of the neutrino probability density may be
compensated by approaching $\zeta\rightarrow -k$ (recall that $F_{i}(\zeta,\upsilon)
\propto (\zeta +k)^{-2}$). This regime is
accompanied by increasing of the neutrino mass $m_{\nu}|_{CLEP}\sim (\zeta +k)^{-1}$.
For a particular value 
$\alpha =\sqrt{3/8}$.
 the cosmological equations allow the following analytic solution for
the late time universe:
$\phi(t)=\frac{M_{p}}{2\alpha}\varphi_{0}+
\frac{M_{p}}{2\alpha}\ln(M_{p}t)$, \,
$a(t)\propto t^{1/3}e^{\lambda t}$,
where $a=a(t)$ is the scale factor,
$\lambda =M_{p}^{-1}\sqrt{\Lambda /3}$ and 
\begin{equation}
\Lambda =
\frac{V_{2}(\upsilon_{clep})+|k|V_{1}(\upsilon_{clep})}{(b-k)^{2}}
\label{Lambda-nu}
\end{equation}
is the cosmological constant of the universe in the CLEP state. 
Such CLEP-neutrino matter is  detectable practically only
through gravitational interaction and this is why it can be regarded
as a model of a dark matter. The mass of CLEP-neutrino increases as 
$a^{3/2}\propto e^{3\lambda t/2}$.
  This dark matter is also cold one in
the sense that kinetic energy of neutrinos is negligible as compared to their
mass. However due to the dynamical fermionic $\Lambda$   term 
generated by neutrinos in CLEP state, this cold dark matter has negative
 pressure and its equation
of state approaches $p_{d.m.}=-\rho_{d.m.}$ as  $a(t)\rightarrow\infty$.
Besides, the energy density of this dark matter scales in a way very
similar to the dark energy which
includes both a cosmological constant and an exponential potential.
Thus, in this toy model, the CLEP dark matter displays itself as a sort of 
a dark energy.

The remarkable feature of such
a Cosmo-Particle solution is that {\em  the total energy density of the
universe
in this case is less than it would be in the universe free
of fermionic matter at all}. In particular, one can check in a simple
model for the Higgs prepotentials $V_{1}$, $V_{2}$ that $\Lambda <\Lambda^{(0)}$.
This means that there are two different
vacua: one is usual vacuum free of the particles, which is
actually a false vacuum, and another one, a true vacuum,
 which could be called "Cosmo-Particle Vacuum" since it is a state
containing neutrinos in CLEP state. Therefore one should expect the possibility
of soft domain walls connecting these two vacua, that may be similar to
soft domain walls studied in Ref.\cite{Schramm}.

One can expect that spherically symmetric solutions in the regime of
the CLEP states
may play an important role in the resolution of the halos dark matter
puzzle.

\end{document}